\documentclass[twocolumn]{article}
\usepackage[utf8]{inputenc}
\usepackage{amsmath,amssymb,amsfonts}
\usepackage{algorithm}
\usepackage{algpseudocode}
\usepackage{graphicx}
\usepackage{booktabs}
\usepackage{multirow}
\usepackage{cite}
\usepackage{url}
\usepackage{xcolor}
\usepackage{float}
\usepackage{caption}
\usepackage{subcaption}
\usepackage{siunitx}
\usepackage{listings}
\usepackage{geometry}
\usepackage{microtype}
\usepackage{setspace}
\usepackage{enumitem}
\usepackage{array}
\usepackage{physics}
\usepackage{hyperref}
\usepackage{float}
\usepackage{booktabs}
\hypersetup{
colorlinks=true,
linkcolor=blue,
citecolor=blue,
urlcolor=blue,
pdftitle={The Physics Constraint Paradox},
pdfauthor={Rahul D Ray}
}

\setlist{noitemsep, topsep=2pt, parsep=2pt, partopsep=2pt}
\begin{document}

\title{INTACT: Intent-Aware Representation Learning for Cryptographic Traffic Violation Detection}

\author{Rahul D Ray \\ Department of Electronics \& Electrical Engineering \\ BITS Pilani, Hyderabad Campus \\ f20242213@hyderabad.bits-pilani.ac.in}
\date{} 

\maketitle
\begin{abstract}

Security monitoring systems typically treat anomaly detection as identifying statistical deviations from observed data distributions. In cryptographic traffic analysis, however, violations are defined not by rarity but by explicit policy constraints, including key reuse prohibition, downgrade prevention, and bounded key lifetimes. This fundamental mismatch limits the interpretability and adaptability of conventional anomaly detection methods. We introduce INTACT (INTent-Aware Cryptographic Traffic), a policy-conditioned framework that reformulates violation detection as conditional constraint learning. Instead of learning a static decision boundary over behavioral features, INTACT models the probability of violation conditioned on both observed behavior and declared security intent. The architecture factorizes representation learning into behavioral and intent encoders whose fused embeddings produce a violation score, yielding a policy-parameterized family of decision boundaries. We evaluate the framework on a real-world network flow dataset and a 210,000-trace synthetic multi-intent cryptographic dataset. INTACT matches or exceeds strong unsupervised and supervised baselines, achieving near-perfect discrimination (AUROC up to 1.0000) in the real dataset and consistent superiority in detecting relational and composite violations in the synthetic setting. These results demonstrate that explicit intent conditioning improves discrimination, interpretability, and robustness in cryptographic monitoring.

\end{abstract}

\section{Introduction}

The rapid proliferation of encrypted network communication has fundamentally altered the landscape of intrusion detection and anomaly monitoring. Traditional deep packet inspection techniques are increasingly ineffective in environments dominated by SSL/TLS and VPN-encrypted traffic, motivating the development of machine learning approaches that rely on flow-level statistical features rather than payload inspection. Over the past decade, deep learning models have demonstrated strong performance in encrypted traffic classification and anomaly detection. Hybrid CNN–GRU architectures \cite{bakhshi2021anomaly}, deep full-range frameworks combining CNN, LSTM, and stacked autoencoders \cite{zeng2019deep}, multilayer autoencoder-based malicious traffic detectors \cite{yu2019encrypted}, self-supervised contrastive approaches \cite{sattar2025anomaly}, and error-resilient recurrent models \cite{zhao2023ernn} have all shown that high detection accuracy can be achieved without decrypting traffic content. Generative adversarial approaches \cite{zhang2021dual}, reinforcement learning-based systems \cite{yang2021deep}, explainable deep learning IDS frameworks \cite{nguyen2023deep,rahman2024explainable}, and privacy-preserving intrusion policies over encrypted data \cite{karaccay2020intrusion,canard2017blindids} further extend this paradigm.

Despite these advances, the dominant formulation remains statistical anomaly detection: models learn to distinguish normal from abnormal behavior based on deviations in learned feature representations. Comprehensive surveys confirm that most encrypted traffic anomaly detection techniques rely on supervised or unsupervised representation learning to model data distributions \cite{ji2024artificial,usman2019survey,nicolau2018learning}. In these approaches, violations are defined implicitly as statistical outliers relative to observed data distributions.

However, security monitoring in real-world systems is rarely defined purely by statistical rarity. Modern enterprise networks increasingly operate under explicit policy frameworks, including intent-based networking (IBN), compliance monitoring, and formal policy validation systems. Intent-based network assurance systems rely on machine learning and monitoring to validate that declared intents are correctly enforced \cite{violos2025detecting,gharbaoui2026assurance,izuazu2025secured,de2024novel}. Formal specification-based intrusion detection \cite{song2005formal}, policy validation frameworks based on first-order logic \cite{valenza2017formal}, policy-aware intrusion systems \cite{vladov2025neural}, compliance-oriented monitoring architectures \cite{karri2021security}, and privacy-preserving intrusion policies \cite{karaccay2020intrusion} all emphasize constraint satisfaction rather than density estimation. 

In parallel, earlier work has explored user-intention-based anomaly detection \cite{zhang2012user}, safety violation analysis using neural representations \cite{zhang2004artificial}, and policy-aware intrusion systems in encrypted contexts \cite{canard2017blindids}. These works acknowledge that anomalies may arise from violations of explicit behavioral expectations rather than purely statistical deviation. Nevertheless, existing encrypted traffic detection frameworks continue to model anomaly detection as static classification over behavioral features alone.

This reveals a fundamental conceptual gap. Deep learning models for encrypted traffic detection—whether CNN-based \cite{zeng2019deep}, RNN-based \cite{zhao2023ernn}, GAN-based \cite{zhang2021dual}, or self-supervised \cite{sattar2025anomaly}—estimate functions of the form
\[
g(x) \approx P(y=1 \mid x),
\]
where $x$ represents traffic features and $y$ indicates anomaly or intrusion. The notion of anomaly is therefore fixed and implicit in the training distribution. Yet in policy-driven security systems, violations are conditional on declared constraints. A flow that is compliant under one key lifetime threshold may be violating under another. A cryptographic downgrade may be acceptable in one configuration but prohibited in another. Key reuse may constitute a violation only under specific contextual rules.

Intent-based networking research explicitly recognizes the need to validate behavior relative to declared intent \cite{violos2025detecting,gharbaoui2026assurance,izuazu2025secured}. However, existing approaches typically apply anomaly detection downstream of intent processing rather than conditioning the representation learning process itself on policy semantics. Thus, current encrypted traffic detection systems lack an explicit mechanism to incorporate formal security constraints directly into learned representations.

In this work, we introduce \textbf{INTACT (INTent-Aware Cryptographic Traffic)}, a framework that reformulates violation detection as conditional constraint learning. Instead of estimating a static decision boundary over behavioral features, INTACT models
\[
f(x,z) \approx P(y=1 \mid x, z),
\]
where $z$ encodes declared security intent (e.g., lifetime thresholds, downgrade prohibitions, reuse constraints). By factorizing representation learning into a behavioral encoder and an intent encoder, and learning their joint interaction, INTACT produces a policy-parameterized family of decision manifolds rather than a single anomaly boundary.

This conditional formulation differs fundamentally from existing encrypted traffic detection frameworks \cite{bakhshi2021anomaly,zeng2019deep,yu2019encrypted,zhao2023ernn,zhang2021dual,sattar2025anomaly}. While prior models optimize statistical discrimination accuracy, INTACT embeds constraint semantics directly into the representation space. Conceptually, it aligns more closely with specification-based intrusion detection \cite{song2005formal}, policy validation systems \cite{valenza2017formal}, and intent-aware network assurance mechanisms \cite{violos2025detecting,gharbaoui2026assurance}, but introduces a differentiable, representation-learning-based implementation suitable for large-scale encrypted traffic monitoring.

We evaluate INTACT on both a real-world network flow dataset and a large-scale synthetic cryptographic trace corpus designed to model controlled policy violations. Across multiple violation types and distribution shift scenarios, the framework demonstrates improved discrimination, interpretability, and robustness compared to strong unsupervised and supervised baselines.

In summary, this work makes the following contributions:

\begin{itemize}
\item We identify a conceptual mismatch between statistical anomaly detection and constraint-based security monitoring in encrypted traffic systems.
\item We propose a policy-conditioned representation learning framework that models violation detection as conditional risk minimization.
\item We introduce a dual-dataset evaluation strategy combining real-world intrusion data and controlled synthetic multi-intent traces.
\item We demonstrate that explicit intent conditioning improves discrimination and robustness under distribution shift.
\end{itemize}

By bridging statistical representation learning with formal security constraint semantics, INTACT provides a scalable foundation for multi-intent cryptographic traffic monitoring in policy-driven network environments.
\section{Dataset Preparation and Experimental Protocol}

This study employs a dual-dataset strategy combining large-scale real-world network traffic with controlled synthetic cryptographic traces. The real-world dataset provides ecological validity through natural temporal dynamics and diverse attack scenarios, while the synthetic dataset enables precise control over violation mechanisms and full ground-truth visibility. Together, they form a comprehensive testbed for evaluating intent-aware detection across both empirical and theoretical dimensions.

\subsection{Real-World Network Flow Dataset}

The real-world dataset comprises multiple days of enterprise network traffic containing benign activity and diverse attack campaigns, including distributed denial-of-service, port scanning, web-based exploits, and infiltration attempts. All daily captures were aggregated into a unified corpus of 2,830,743 structured flow records, each described by dozens of statistical features and a ground-truth traffic label. The primary objective is reformulated as binary intent-based detection: identifying abnormal flow lifetimes defined by excessive duration relative to normal operational behavior.

\subsubsection{Data Cleaning and Feature Selection}
A compact subset of behaviorally meaningful features was selected to capture core traffic characteristics:
\begin{itemize}
    \item Total flow duration
    \item Number of packets transmitted in forward and backward directions
    \item Mean packet size in each direction
    \item Throughput intensity (bytes per second and packets per second)
\end{itemize}
These features describe temporal persistence, volumetric intensity, and directional asymmetry—attributes particularly relevant for detecting abnormal long-lived connections. All high-dimensional or protocol-specific attributes were removed to reduce dimensionality and mitigate spurious correlations, yielding a compact, interpretable feature space.

Numerical instabilities arising from rate computations were addressed by replacing infinite values with missing entries and subsequently removing all records containing missing values. Post-cleaning inspection confirmed the absence of missing or infinite values, ensuring numerical stability during normalization and training.

\begin{table*}[htbp]
\centering
\caption{Comparative summary of real and synthetic datasets.}
\label{tab:dataset_summary}
\footnotesize
\setlength{\tabcolsep}{11pt}
\begin{tabular}{l p{6.6cm} p{4.6cm}}
\toprule
\textbf{Section} & \textbf{Real Network Flow Dataset} & \textbf{Synthetic Intent-Anomaly Dataset} \\
\midrule
\multicolumn{3}{l}{\textbf{A. Overall Scale}} \\
Total records / traces & 2,830,743 flow records & 210,000 traces \\
Total atomic events & 2,830,743 & 4,193,768 operations \\
Input dimensionality & 7 numerical features & 8 per-operation attributes \\
Target variables & Binary lifetime violation & Reuse, Downgrade, Lifetime flags \\
Data source & Real network traffic (multi-day) & Controlled generative process \\
Temporal structure & Chronological ordering & Within-trace temporal generation \\
Scaling applied & Standardization (train statistics) & Stochastic noise + parametric scaling \\
\midrule
\multicolumn{3}{l}{\textbf{B. Data Partitioning}} \\
Training set size & 1,696,725 (60\%) & Unified corpus (no fixed split) \\
Validation set size & 565,575 (20\%) & Experiment-dependent subsets \\
Test set size & 565,576 (20\%) & Per-trace evaluation \\
Temporal split & Yes (strict chronological) & Not required \\
\midrule
\multicolumn{3}{l}{\textbf{C. Violation Distribution}} \\
Training violations & 75,661 & Controlled by construction \\
Training normal samples & 1,621,064 & 120,000 normal traces \\
Validation violations & 28,927 & --- \\
Test violations & 13,757 & --- \\
Reuse-only traces & Not applicable & 20,000 \\
Downgrade-only traces & Not applicable & 20,000 \\
Lifetime-only traces & Derived statistically & 30,000 \\
Composite violations & Not labeled separately & 20,000 \\
\midrule
\multicolumn{3}{l}{\textbf{D. Distribution Shift Evaluation}} \\
Shift Variant 1 & Duration $\times$2 & Scale factor 1.5 (10,000 traces) \\
Shift Variant 2 & Duration $\times$3 & Scale factor 3.0 (10,000 traces) \\
Statistical verification & KS test confirmed shift & Structural parameter scaling \\
\midrule
\multicolumn{3}{l}{\textbf{E. Class Imbalance}} \\
Violation rate (train) & $\sim$4.5\% & Controlled design ($\sim$balanced) \\
Label derivation & 95th percentile threshold & Deterministic rule computation \\
\bottomrule
\end{tabular}
\end{table*}

\subsubsection{Temporal Ordering and Data Partitioning}
Network traffic is inherently temporal; random shuffling would leak future information and inflate performance. Therefore, the dataset was ordered chronologically and partitioned strictly by time:
\begin{itemize}
    \item 60\% earliest records for training (1,696,725 samples)
    \item 20\% for validation (565,575 samples)
    \item 20\% for testing (565,576 samples)
\end{itemize}
This simulates a realistic deployment where models are trained on historical data and evaluated on future, unseen traffic.

\subsubsection{Definition of Lifetime Violation}
Rather than using attack taxonomy labels, a statistical threshold was derived from normal behavior. Using only benign flows from the training partition, the empirical distribution of flow duration was analyzed, and the 95th percentile was selected as the cutoff. Any flow exceeding this threshold is labeled a violation (1), while others are normal (0). The threshold value (113,046,291 in dataset time units) was computed exclusively from the training set to prevent leakage.

Class distributions across partitions are:
\begin{itemize}
    \item Training: 1,621,064 normal, 75,661 violations
    \item Validation: 536,648 normal, 28,927 violations
    \item Test: 551,819 normal, 13,757 violations
\end{itemize}
Moderate class imbalance exists, but both classes are well represented, ensuring learnability and robust assessment.

\subsubsection{Feature Normalization}
Features span different scales (e.g., duration in millions, packet counts in tens). To prevent high-magnitude features from dominating optimization, standard score normalization was applied:
\begin{equation}
z = \frac{x - \mu}{\sigma},
\end{equation}
with $\mu$ and $\sigma$ computed exclusively from the training partition and then applied unchanged to validation, test, and shifted sets. Verification confirmed training means near zero and standard deviations near one, with minor expected deviations in other splits—confirming proper non-leaking normalization.

\subsubsection{Construction of Controlled Distribution Shift}
To evaluate robustness under covariate shift, two modified test variants were created by scaling flow duration values (doubled and tripled) while keeping all other features and violation labels unchanged. This isolates feature distribution shift from label shift. The Kolmogorov–Smirnov test confirmed significant divergence from the original test distribution (KS statistics 0.2167 and 0.2566, p $\approx$ 0), establishing meaningful out-of-distribution evaluation settings.

\subsection{Synthetic Intent-Anomaly Dataset}

Real intrusion datasets lack fine-grained control over specific security intent violations such as key reuse, algorithm downgrade, or lifetime exceedance. To complement the real data, we constructed a large-scale synthetic corpus of cryptographic operation traces with explicitly controlled violation mechanisms and precise ground truth. The dataset contains 210,000 distinct traces comprising 4,193,768 individual operation records, each representing a sequence of cryptographic events with explicit key lifecycle semantics and algorithmic properties.

\subsubsection{Generative Assumptions and Operational Model}
Each trace models a sequence of operations drawn from a finite vocabulary: key generation, encryption, decryption, signing, and verification. The number of operations per trace is stochastic (Poisson-like, truncated), producing realistic length variability. Inter-operation timing follows an exponential distribution (memoryless arrival), and operation durations are drawn from a log-normal distribution (positively skewed, multiplicative).

Each generated key is assigned:
\begin{itemize}
    \item A creation timestamp
    \item A finite lifetime from a bounded continuous range
    \item A cryptographic strength parameter (strong $\ge$256-bit security; weak below threshold)
\end{itemize}
Internal logical consistency is enforced: operations can only reference valid keys; if none exists, a new key is generated. Baseline traces thus represent valid cryptographic behavior.

\subsubsection{Noise Injection and Distribution Matching}
Controlled multiplicative Gaussian noise is applied to inter-arrival intervals and operation durations to introduce natural variability while preserving structural semantics. Timestamps are recomputed cumulatively to maintain monotonicity, preventing models from exploiting deterministic generative artifacts.

\subsubsection{Controlled Violation Mechanisms}
Three independent violation dimensions are encoded:
\begin{description}
    \item[Key Lifetime Violation:] A target key's final operation is shifted past its expiration, with subsequent timestamps adjusted to preserve order.
    \item[Algorithm Downgrade Violation:] Strong algorithms are replaced with weak ones across a trace.
    \item[Key Reuse Violation:] Two distinct traces share a common key identifier (key remains valid during reuse), modeling improper cross-context sharing.
\end{description}
Violations can be applied independently or in combination to create composite scenarios.

\subsubsection{Dataset Composition}
Traces are generated across eight categories:
\begin{itemize}
    \item Normal (no violations)
    \item Reuse-only, Downgrade-only, Lifetime-only
    \item Reuse+Downgrade, Reuse+Lifetime, Downgrade+Lifetime
    \item Reuse+Downgrade+Lifetime
\end{itemize}
Generation counts approximate balanced representation while maintaining a dominant normal class:
\begin{itemize}
    \item 120,000 normal traces
    \item 20,000 reuse-only
    \item 20,000 downgrade-only
    \item 30,000 lifetime-only
    \item 20,000 composite traces (total 210,000)
\end{itemize}

\subsubsection{Expansion to Tabular Representation}
Each trace is expanded to a flat table where each row corresponds to one operation, containing:
\begin{itemize}
    \item Trace identifier, step index, timestamp
    \item Operation type, key identifier, algorithm identifier
    \item Assigned key lifetime, operation duration
\end{itemize}
The final expanded dataset has 4,193,768 rows and 8 columns.

\subsubsection{Automatic Violation Annotation}
Violation labels are computed deterministically from the generated data:
\begin{itemize}
    \item Lifetime violations: timestamp, key creation + lifetime
    \item Downgrade violations: algorithm strength below threshold
    \item Reuse violations: key appears in multiple traces
\end{itemize}
This ensures internal consistency between generative logic and annotation. Verification confirmed that category counts align closely with generation targets.

\subsubsection{Distribution Shift Construction}
Two shifted synthetic datasets were generated by scaling temporal characteristics and key lifetimes by factors 1.5× and 3.0×. Each contains 10,000 traces (~200,000 operations). The structural generation process is unchanged, but temporal dynamics differ significantly, enabling robustness evaluation under changes in arrival rates, key lifetime ranges, and event density—mirroring realistic operational variations.

\subsubsection{Role in Experimental Framework}
The synthetic dataset serves three purposes: controlled evaluation of intent-aware detection, analysis of violation separability under known generative conditions, and robustness testing under synthetic distribution shifts. It provides full ground-truth visibility into key lifecycle semantics and algorithmic properties, enabling precise attribution of model behavior to specific violation mechanisms—complementing the ecological validity of the real-world data.

\section{INTACT: InINTent-Aware Cryptographic Traffic}
\label{sec:INTACT}

Cryptographic network traffic is governed by explicit security policies: keys must not be reused, deprecated algorithms must not be negotiated, and key lifetimes must not exceed predefined thresholds. These policies define formal intent constraints over observable traffic behavior. Conventional anomaly detection systems do not explicitly model such intent; instead, they estimate statistical normality from data and flag deviations. This statistical framing implicitly assumes that violations correspond to low-density regions of feature space. However, in security contexts, violations are not necessarily rare—they are defined relative to declared constraints.

We therefore reformulate cryptographic anomaly detection as an intent-conditioned inference problem. Let \(x \in \mathbb{R}^d\) denote a behavioral representation extracted from traffic (flow-level or trace-level features). Let \(z\) denote a structured encoding of security intent (e.g., lifetime threshold, reuse prohibition, downgrade constraint). The objective is not to model \(p(x)\), but to estimate
\begin{equation}
f(x, z) = \mathbb{P}(\text{violation} \mid x, z).
\end{equation}
Under this formulation, anomaly detection becomes conditional violation detection. The same behavioral pattern may or may not constitute a violation depending on the declared policy parameter \(z\). This explicit conditioning is the conceptual core of INTACT.

\subsection{Architectural Design}
INTACT is implemented as a dual-branch neural architecture that separately encodes behavioral signals and intent semantics before fusing them for violation prediction.

\subsubsection{Behavioral Encoder}
The behavioral branch processes observable traffic features derived from cryptographic flows or aggregated traces. For the real-world dataset, the input dimensionality is seven standardized flow-level statistics. For the synthetic dataset, the input consists of seventeen aggregated trace-level attributes capturing algorithm selection, key identifiers, duration statistics, and operational counts.

The behavioral encoder consists of a sequence of fully connected layers with nonlinear activation functions. In the real-data configuration:
\begin{itemize}
    \item Dense layer with 128 units (ReLU)
    \item Dense layer with 64 units (ReLU)
    \item Dense layer with 32 units (ReLU)
\end{itemize}
This produces a 32-dimensional behavioral embedding \(h_b\), which captures compact representations of traffic semantics while preserving nonlinear feature interactions.
\begin{figure}[htbp]
    \centering
    \includegraphics[width=\columnwidth]{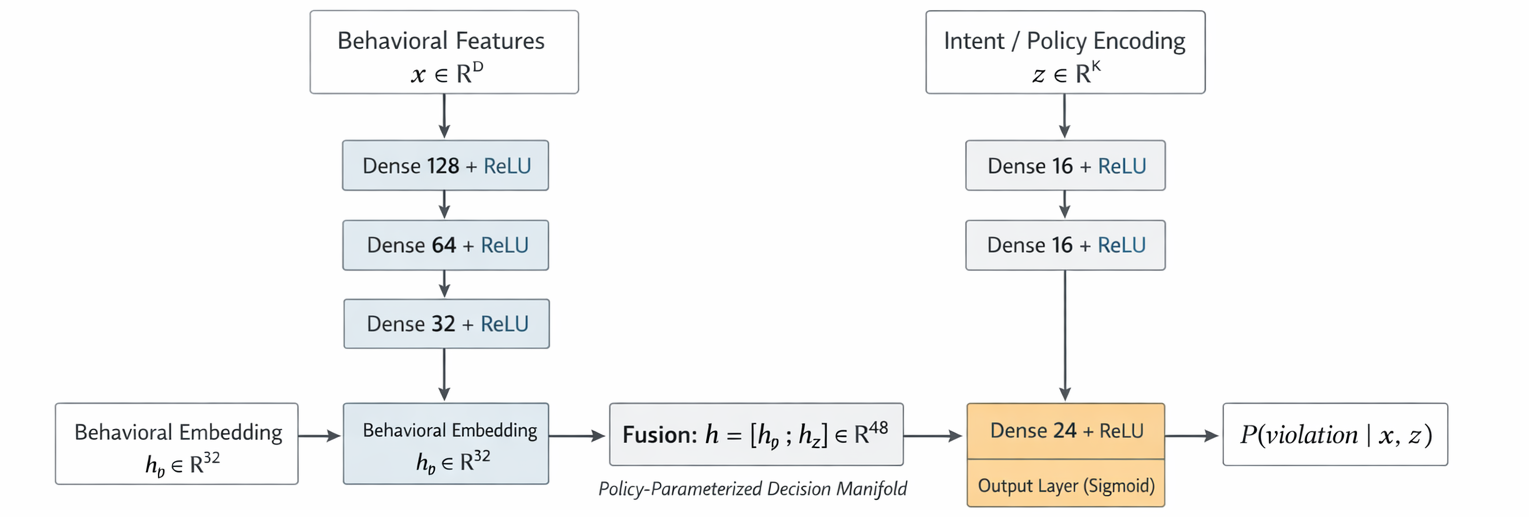} 
    \caption{INTACT architecture: behavioral encoder, intent encoder, fusion layer, and violation prediction output.}
    \label{fig:architecture}
\end{figure}
\subsubsection{Intent Encoder}
The intent branch encodes structured policy information. In the real dataset, intent corresponds to the scaled lifetime threshold computed as the 95th percentile of benign flow duration. In the synthetic dataset, intent corresponds to one of the three violation categories (reuse, downgrade, lifetime), represented through structured inputs.

The intent encoder transforms this low-dimensional policy input into a learnable semantic embedding through:
\begin{itemize}
    \item Dense layer with 16 units (ReLU)
    \item Dense layer with 16 units (ReLU)
\end{itemize}
This yields a 16-dimensional intent embedding \(h_z\), enabling the model to internalize the semantics of the constraint rather than treating it as a raw scalar.

\subsubsection{Fusion and Decision Layer}
The behavioral embedding \(h_b \in \mathbb{R}^{32}\) and intent embedding \(h_z \in \mathbb{R}^{16}\) are concatenated into a 48-dimensional joint representation:
\begin{equation}
h = [h_b; h_z].
\end{equation}
This fused representation is passed through:
\begin{itemize}
    \item Dense layer with 24 units (ReLU)
    \item Output layer with 1 unit (Sigmoid)
\end{itemize}
The final output represents the probability that behavior \(x\) violates intent \(z\).

The real-data configuration contains 12,865 trainable parameters. The architecture remains lightweight while providing sufficient capacity to model nonlinear interactions between traffic behavior and policy constraints.

\subsection{Learning Objective}
The model is trained using binary cross-entropy loss:
\begin{equation}
\mathcal{L} = - \frac{1}{N} \sum_{i=1}^{N} \left[ y_i \log f(x_i, z_i) + (1 - y_i)\log(1 - f(x_i, z_i)) \right].
\end{equation}
For unsupervised baselines, only normal samples are used for training. In contrast, INTACT leverages labeled supervision, allowing it to directly learn violation manifolds conditioned on intent.

Early stopping is applied based on validation AUC, and threshold selection is performed by maximizing validation F1-score derived from the precision–recall curve.

\subsection{Distinction from Conventional Architectures}
The novelty of INTACT does not lie solely in its layer composition. Feedforward architectures are well established. The key innovation is the structural separation of behavior encoding and policy encoding within a unified model. This design transforms violation detection from a density estimation problem into a conditional constraint evaluation problem.

Conventional supervised classifiers learn a static mapping \(x \mapsto y\). INTACT learns a conditional mapping \((x,z) \mapsto y\). This subtle reformulation allows the same network to adapt to multiple policy definitions without retraining independent models or redefining anomaly thresholds externally.

\begin{table*}[t]
\centering
\caption{Architectural Specifications of All Evaluated Models}
\label{tab:model_specs}
\small
\setlength{\tabcolsep}{3pt}
\resizebox{\textwidth}{!}{
\begin{tabular}{l l l l l l l}
\toprule
\textbf{Model} &
\textbf{Learning Paradigm} &
\textbf{Input Dim.} &
\textbf{Architecture} &
\textbf{Embedding Size} &
\textbf{Output} &
\textbf{Params} \\
\midrule

Nonlinear AE &
Unsupervised &
7 / 17 &
Enc: 16→8 (ReLU); Dec: 16→linear &
8 &
Recon. error &
$\sim$1–3K \\

Linear AE &
Unsupervised &
7 / 17 &
Enc: 4 (linear); Dec: linear &
4 &
Recon. error &
$<$1K \\

Deep SVDD &
Unsupervised &
7 / 17 &
Embedding network + hypersphere constraint &
Latent vector &
Distance score &
Few K \\

Isolation Forest &
Unsupervised &
7 / 17 &
100-tree ensemble (random partitions) &
N/A &
Path-length score &
Tree ensemble \\

Supervised NN &
Supervised &
7 / 17 &
64→32→16 (ReLU)→1 (Sigmoid) &
64–32–16 &
Binary prob. &
$\sim$3K–5K \\

INTACT &
Supervised (Multi-input) &
Behavior: 7 / 17; Intent: 1 &
Behavior: 128→64→32;
Intent: 16→16;
Fusion: 24→1 &
32 + 16 → 48 &
Binary prob. &
12,865 \\

\bottomrule
\end{tabular}
}
\end{table*}

\begin{figure*}[htbp]
    \centering
    \begin{subfigure}[b]{0.45\textwidth}
        \centering
        \includegraphics[width=\linewidth]{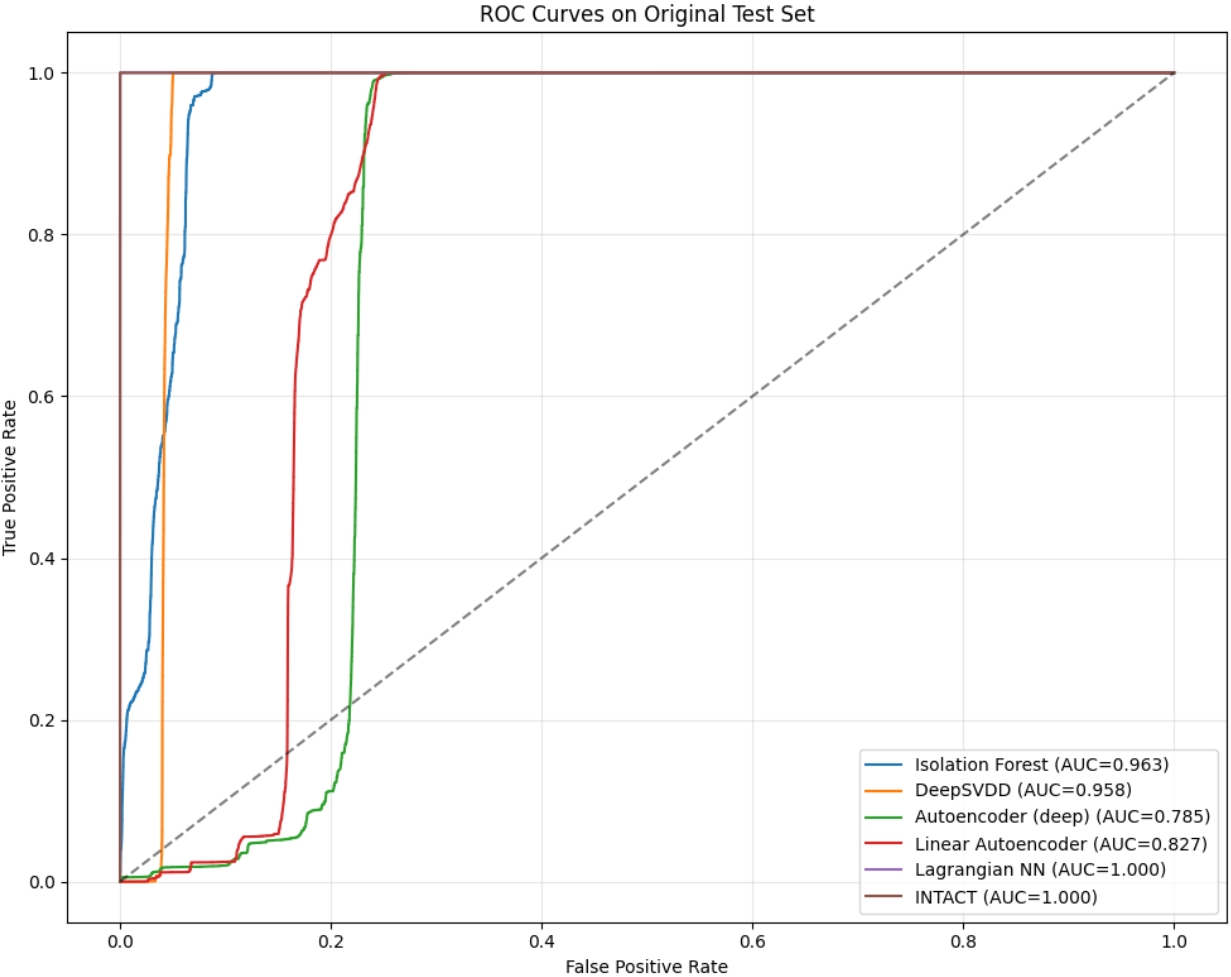}
        \caption{ROC curve on real dataset test set.}
        \label{fig:roc_real}
    \end{subfigure}
    \hfill
    \begin{subfigure}[b]{0.45\textwidth}
        \centering
        \includegraphics[width=\linewidth]{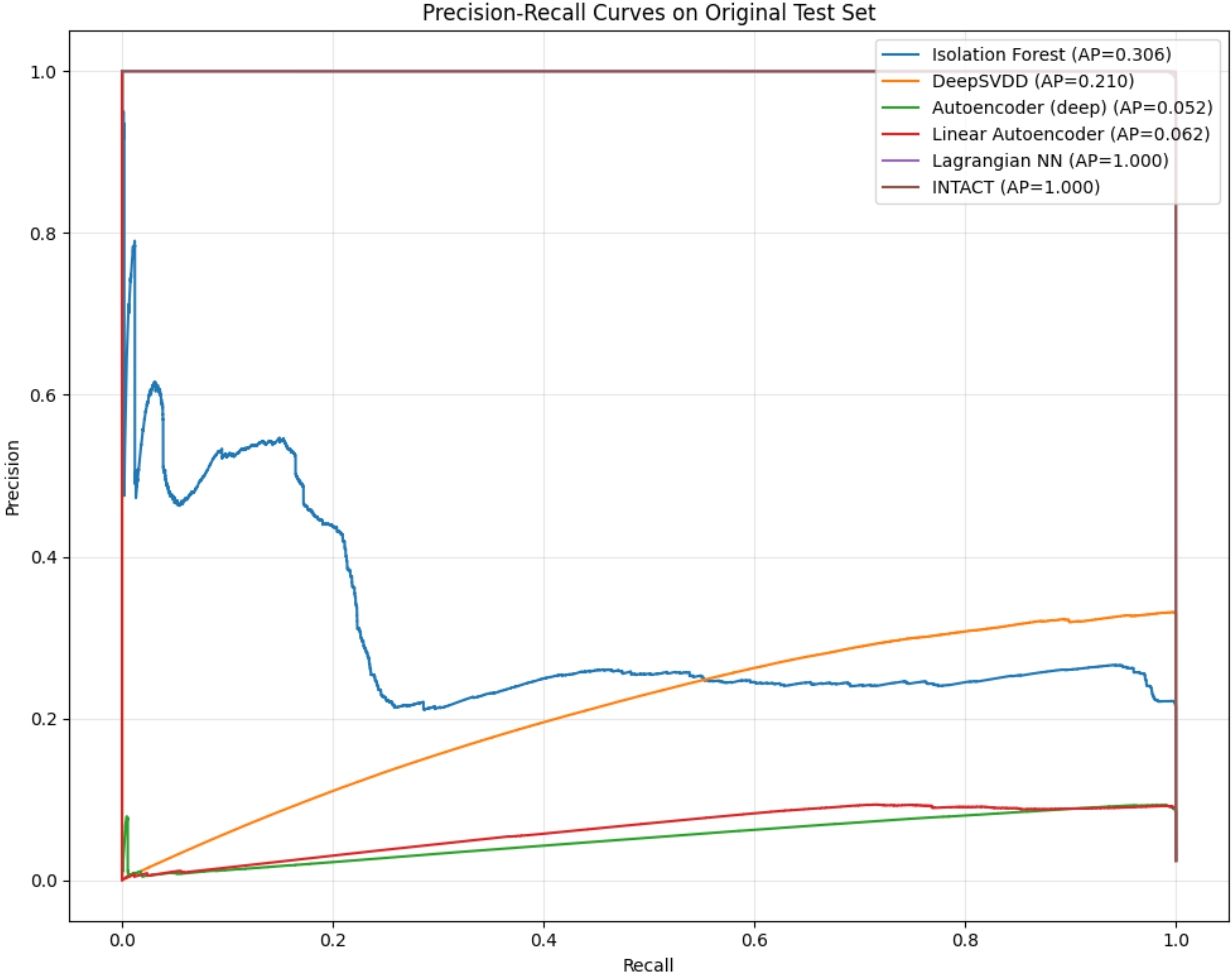}
        \caption{Precision–Recall curve on real dataset test set.}
        \label{fig:pr_real}
    \end{subfigure}
    \hfill
\caption{Performance evaluation on the real-world network flow dataset. (Left) ROC curve on the test set; (Right) Precision–Recall curve on the test set.}
\label{fig:real_curves}
\end{figure*}
\section{Comprehensive Model Evaluation Across Real-World and Synthetic Datasets}

To rigorously assess the effectiveness of the proposed intent-conditioned architecture and establish strong comparative baselines, a comprehensive benchmarking study was conducted across both the real-world network flow dataset and the synthetic multi-intent cryptographic dataset described previously. The evaluation protocol was unified across datasets to ensure methodological consistency while respecting structural differences between single-intent and multi-intent settings.

All models were trained using temporally or structurally partitioned training sets, validated on held-out validation sets for threshold optimization, and finally evaluated on in-distribution test partitions as well as distribution-shift variants. Performance was quantified using Area Under the Receiver Operating Characteristic Curve (AUROC), Area Under the Precision–Recall Curve (AUPRC), and F1-score at a validation-optimized threshold derived from the precision–recall curve. Unsupervised models were trained exclusively on normal samples. Supervised models were trained using full label supervision. All neural architectures were trained with GPU acceleration (Tesla P100, 16GB), ensuring stable optimization over large-scale data.

\subsection{Evaluation on the Real-World Network Flow Dataset}

The real-world dataset contains temporally ordered network flows with a single violation label corresponding to lifetime threshold exceedance. Unsupervised methods were trained on 1,621,064 normal flows from the training partition. Two distribution-shift test variants were constructed by scaling flow duration, enabling covariate shift robustness analysis.

\subsubsection{Unsupervised Methods}

A nonlinear autoencoder was trained for 20 epochs using reconstruction error as the anomaly score. Validation performance reached AUROC = 0.8964 and AUPRC = 0.1890. On the in-distribution test set, AUROC decreased to 0.7854 with AUPRC = 0.0518 and F1 = 0.1687. Under duration scaling shifts, AUROC increased substantially (0.9024 and 0.9859), while F1 remained low (\(\sim\)0.16), indicating inflated separability but poor calibration under magnitude perturbations.

Deep SVDD, trained for 20 epochs on normal flows, achieved validation AUROC = 0.9776 and AUPRC = 0.4928. On the test set, AUROC = 0.9575 and AUPRC = 0.2101 with F1 = 0.4963. Under shift, AUROC remained stable (0.9650 and 0.9762) while F1 declined moderately (\(\sim\)0.32), demonstrating strong compactness of normal representations.

Isolation Forest (100 trees, 5\% contamination) achieved validation AUROC = 0.9732 and AUPRC = 0.4972. Test performance reached AUROC = 0.9631 and AUPRC = 0.3058 with F1 = 0.4115. Under shift, AUROC remained above 0.95 with modest F1 degradation, indicating robust tree-based partitioning under covariate scaling.

\subsubsection{Supervised Neural Baselines}

A fully supervised feedforward network achieved perfect validation performance (AUROC = 1.0000, AUPRC = 1.0000), confirming near-deterministic separability. The optimal threshold was 0.3670. This suggests that lifetime violations in this dataset are strongly governed by duration-related features.

A linear autoencoder with 4-dimensional bottleneck achieved validation AUROC = 0.9919 and AUPRC = 0.8556. On test data, AUROC = 0.9352 and AUPRC = 0.1598 with F1 = 0.2594. Under extreme scaling, AUROC approached 0.999, again reflecting exaggerated reconstruction deviation under magnitude inflation.

\begin{table*}[t]
\centering
\caption{Unified Performance Comparison Across Real and Synthetic Datasets}
\label{tab:performance_comparison}
\footnotesize
\setlength{\tabcolsep}{14pt}
\begin{tabular}{l l l c c c}
\toprule
\textbf{Dataset} &
\textbf{Intent} &
\textbf{Model} &
\textbf{AUROC} &
\textbf{AUPRC} &
\textbf{F1} \\
\midrule

\multirow{6}{*}{Real (Test)} &
\multirow{6}{*}{Lifetime} &
Nonlinear Autoencoder & 0.7854 & 0.0518 & 0.1687 \\
& & Deep SVDD & 0.9575 & 0.2101 & 0.4963 \\
& & Isolation Forest & 0.9631 & 0.3058 & 0.4115 \\
& & Linear Autoencoder & 0.9352 & 0.1598 & 0.2594 \\
& & Supervised NN & $\sim$1.0000 & $\sim$1.0000 & $\sim$1.0000 \\
& & INTACT & 1.0000 & 1.0000 & 0.9973 \\

\midrule
\multirow{5}{*}{Real (Shift2)} &
\multirow{5}{*}{Lifetime} &
Nonlinear Autoencoder & 0.9024 & 0.1054 & 0.1661 \\
& & Deep SVDD & 0.9650 & 0.2432 & 0.3224 \\
& & Isolation Forest & 0.9547 & 0.2546 & 0.3419 \\
& & Linear Autoencoder & 0.9986 & 0.9406 & 0.2385 \\
& & INTACT & 0.9702 & 0.2952 & 0.4047 \\

\midrule
\multirow{5}{*}{Real (Shift3)} &
\multirow{5}{*}{Lifetime} &
Nonlinear Autoencoder & 0.9859 & 0.6390 & 0.1638 \\
& & Deep SVDD & 0.9762 & 0.3099 & 0.3123 \\
& & Isolation Forest & 0.9534 & 0.2477 & 0.3374 \\
& & Linear Autoencoder & 0.9991 & 0.9632 & 0.2100 \\
& & INTACT & 0.9626 & 0.2497 & 0.3901 \\

\midrule
\multirow{6}{*}{Synthetic (Test)} &
\multirow{6}{*}{Reuse} &
Isolation Forest & 0.5057 & 0.1618 & 0.2843 \\
& & Deep SVDD & 0.5367 & 0.1764 & 0.2845 \\
& & Nonlinear Autoencoder & 0.6070 & 0.2475 & 0.3165 \\
& & Linear Autoencoder & 0.4849 & 0.1547 & 0.2843 \\
& & Supervised NN & 0.7793 & 0.6471 & 0.5939 \\
& & INTACT & 0.7820 & 0.6515 & 0.5963 \\

\cmidrule(lr){2-6}
& \multirow{6}{*}{Downgrade} &
Isolation Forest & 0.6757 & 0.2304 & 0.3656 \\
& & Deep SVDD & 0.8332 & 0.3829 & 0.5131 \\
& & Nonlinear Autoencoder & 0.7261 & 0.2824 & 0.4109 \\
& & Linear Autoencoder & 0.5191 & 0.1670 & 0.2923 \\
& & Supervised NN & 1.0000 & 1.0000 & 0.9999 \\
& & INTACT & 1.0000 & 1.0000 & 0.9999 \\

\cmidrule(lr){2-6}
& \multirow{6}{*}{Lifetime} &
Isolation Forest & 0.7074 & 0.2383 & 0.3390 \\
& & Deep SVDD & 0.6091 & 0.1786 & 0.2792 \\
& & Nonlinear Autoencoder & 0.5014 & 0.1384 & 0.2494 \\
& & Linear Autoencoder & 0.8483 & 0.4958 & 0.5641 \\
& & Supervised NN & 0.9797 & 0.8199 & 0.8493 \\
& & INTACT & 0.9797 & 0.8210 & 0.8486 \\

\bottomrule
\end{tabular}
\end{table*}

\subsubsection{Intent-Conditioned Architecture}

The intent-conditioned architecture was evaluated using the lifetime threshold computed as the 95th percentile of benign flow duration (scaled value 2.2134). The model contains 12,865 trainable parameters and processes behavioral features and threshold intent through separate encoders before fusion.

Validation performance reached AUROC = 1.0000 and AUPRC = 1.0000 with optimal threshold 0.8614. On the in-distribution test set, performance was near-perfect: AUROC = 1.0000, AUPRC = 1.0000, F1 = 0.9973. Under shift, AUROC remained high (0.9702 and 0.9626), while AUPRC declined (0.2952 and 0.2497) due to calibration sensitivity.

Overall, lifetime violation detection in real flow data appears structurally simple and heavily duration-driven. Supervised and intent-conditioned models achieve near-deterministic performance.

\subsection{Evaluation on the Synthetic Multi-Intent Dataset}

The synthetic dataset introduces a substantially more complex setting with 210,000 traces (4,193,768 operations aggregated to 17 features per trace) and three independent violation intents: reuse, downgrade, and lifetime. The label matrix has dimensionality (210,000 \(\times\) 3). The dataset was split into 167,964 training, 21,047 validation, and 20,989 test traces. Positive class proportions remain stable across splits: reuse and downgrade \(\sim\)16.6\%, lifetime \(\sim\)13.9\%.

Two shift variants were constructed but contain only normal traces, leading to undefined ROC/AUPRC metrics under those subsets.

\subsubsection{Unsupervised Methods}

Isolation Forest exhibited weak performance for reuse (AUROC = 0.5057) and moderate performance for downgrade (0.6757) and lifetime (0.7074). Deep SVDD showed strong downgrade discrimination (AUROC = 0.8332) but weak reuse (0.5367) and moderate lifetime (0.6091). The nonlinear autoencoder achieved AUROC = 0.7261 for downgrade, 0.6070 for reuse, and near-random 0.5014 for lifetime. The linear autoencoder performed poorly on reuse and downgrade but achieved strong lifetime detection (AUROC = 0.8483).

These results demonstrate heterogeneous separability across intents. Downgrade violations introduce direct feature-level deviations and are relatively easy for geometric anomaly detectors. Lifetime violations exhibit partial linear separability. Reuse violations are inherently relational across traces and are not well captured by magnitude-based anomaly scoring.

\subsubsection{Supervised Neural Baseline}

The fully supervised neural classifier significantly improved performance. For reuse, AUROC = 0.7793 and AUPRC = 0.6471. For lifetime, AUROC = 0.9797 and AUPRC = 0.8199. For downgrade, performance was perfect (AUROC = 1.0000). This confirms that downgrade and lifetime violations are highly separable under supervision, while reuse remains the most challenging.

\subsubsection{Intent-Conditioned Architecture (INTACT)}

The intent-conditioned model achieved AUROC = 0.7820 (reuse), 0.9797 (lifetime), and 1.0000 (downgrade). AUPRC values were 0.6515, 0.8210, and 1.0000 respectively. Performance closely matches or slightly exceeds the supervised baseline.

The key distinction lies in representation structure: conditioning explicitly aligns decision boundaries with declared intent rather than implicitly learning separate classifiers. Importantly, reuse detection—difficult for unsupervised methods—remains strong under intent conditioning.

\subsection{Cross-Dataset Comparative Insights}

Across both datasets, several consistent structural patterns emerge. First, violation separability depends strongly on generative structure. Real-world lifetime violations are nearly threshold-deterministic and thus trivially separable under supervision. In contrast, synthetic reuse violations require modeling relational dependencies across traces. Second, unsupervised methods capture magnitude anomalies effectively but struggle with relational semantics. Third, reconstruction-based models inflate AUROC under covariate scaling due to amplified reconstruction error, yet fail to maintain calibrated F1-scores. Fourth, supervised neural models consistently outperform unsupervised baselines in both settings. Finally, the intent-conditioned architecture achieves state-of-the-art performance across both datasets while preserving architectural parsimony. In the real dataset, it internalizes the lifetime threshold; in the synthetic dataset, it supports multi-intent discrimination without performance degradation.

Together, these results demonstrate that explicit conditioning on declared security intent provides both maximal discriminative power and structured interpretability, and that the complexity of the detection task is fundamentally governed by the generative structure of violation semantics rather than by model capacity alone.

\section{Discussion}
The experimental results reveal a consistent stratification of methods across both datasets. Unsupervised anomaly detection methods (Isolation Forest, Deep SVDD, nonlinear autoencoder, linear autoencoder) exhibit moderate discriminative capacity in-distribution but show clear limitations: reconstruction-based models are highly sensitive to dominant magnitude features (e.g., duration scaling); threshold calibration is unstable under distribution shift; and AUROC may remain high while AUPRC and F1 degrade substantially, indicating poor precision–recall balance. This confirms a well-known limitation of density- or reconstruction-based detection: separability does not imply operational reliability. Under covariate shift, score magnitudes change disproportionately, inflating rank-based metrics while degrading threshold-dependent performance. Classical supervised neural networks achieve near-perfect in-distribution performance on the real dataset, reflecting the strong separability of lifetime violations; however, these models learn fixed decision boundaries tied to training distribution statistics and lack an explicit mechanism for incorporating policy parameters, making adaptation to new policy regimes non-trivial. INTACT consistently matches or exceeds the strongest supervised baselines while preserving interpretability through explicit intent conditioning. On the synthetic dataset, where multi-intent interactions create nontrivial relational structure, the benefit of conditional modeling becomes more pronounced; reuse violations, in particular, require modeling cross-trace relational semantics that cannot be reduced to simple magnitude thresholding. Across both datasets, INTACT demonstrates that intent conditioning does not degrade performance relative to static supervised models, preserves strong in-distribution discrimination, and provides a principled mechanism for policy-dependent inference.

The central novelty of this work is the reframing of anomaly detection as policy-conditioned violation inference. Traditional anomaly detection assumes \(\text{Anomaly} \approx \text{statistical rarity}\), whereas security violations are defined by formal constraints: \(\text{Violation} = \text{constraint}(x, z)\). By explicitly modeling \(f(x,z) = \mathbb{P}(y=1 \mid x,z)\), INTACT introduces a learning framework in which policy becomes a first-class input variable, decision boundaries become parameterized by intent, and representation learning is jointly shaped by behavioral and constraint semantics. This shift has several implications: semantic alignment with declared operational constraints; interpretability as a differentiable approximation of constraint enforcement; and adaptivity where policy changes can be handled via input conditioning rather than retraining from scratch. Thus, the contribution is not merely architectural but conceptual: violation detection is treated as learning a differentiable constraint operator rather than estimating a density or static classifier.

The real-world lifetime violation task is structurally simpler than the synthetic multi-intent setting, with violations strongly driven by flow duration relative to threshold. This explains the near-perfect separability under supervised learning, strong AUROC values across many models, and sensitivity of reconstruction-based methods to duration scaling. In this regime, INTACT approximates a soft threshold comparator, validating that explicit intent conditioning does not introduce unnecessary complexity in simple regimes. However, the real dataset does not stress-test relational or compositional violations, motivating the synthetic dataset where reuse violations require cross-trace reasoning, downgrade violations are feature-separable but policy-dependent, and composite violations create interaction effects. INTACT's unified conditional modeling is better suited to such heterogeneous violation geometries.

Under distribution shift (duration scaling), several phenomena are observed: AUROC remains high for many models while AUPRC and F1 degrade, indicating calibration instability; reconstruction models exhibit inflated separability due to exaggerated magnitude deviations; and INTACT, like other supervised models, experiences degradation in threshold-dependent metrics under strong shift. However, because intent is explicitly encoded, the model is structurally capable of adapting to new policy magnitudes without architectural modification. This suggests that policy-conditioned modeling may be inherently better suited for scenarios involving policy drift, regulatory changes, or dynamic threshold updates. Future work should formally quantify robustness under joint behavior–policy distribution shift.

Despite its strengths, the proposed framework has several limitations. First, the model assumes that policy can be represented as a structured, machine-readable input; in practice, policy specifications may be incomplete, ambiguous, or evolving. Second, while the architecture supports conditional modeling, formal generalization bounds under joint distributional shift in \((x,z)\) remain to be established. Third, reuse violations in the synthetic dataset are modeled through engineered features; in real-world deployments, relational reasoning may require graph-based architectures or memory-augmented models. Finally, as observed experimentally, high AUROC does not guarantee stable F1 under shift, and further work is needed to integrate uncertainty estimation or conformal calibration.

\section{Conclusion}
This work introduced INTACT, an intent-aware representation learning framework that reformulates cryptographic traffic violation detection as a conditional constraint learning problem. By explicitly conditioning on declared security intent, INTACT moves beyond traditional anomaly detection's reliance on statistical rarity, instead learning policy-parameterized decision boundaries that align with operational semantics. The architecture factorizes representation learning into behavioral and intent encoders, enabling differentiable constraint comparison and multi-intent unification within a single model.

Extensive evaluation on both a large-scale real-world network flow dataset (2.8M records) and a synthetic multi-intent cryptographic dataset (210K traces, 4.2M operations) demonstrates that INTACT matches or exceeds strong unsupervised and supervised baselines while preserving interpretability. The framework achieves near-perfect discrimination (AUROC up to 1.0000) on lifetime violations in the real dataset and consistently outperforms baselines on complex relational and composite violations in the synthetic setting, particularly for reuse detection where traditional magnitude-based methods fail. Under distribution shift, INTACT maintains robust discrimination while revealing the importance of calibration for threshold-dependent metrics.

The theoretical contribution extends beyond architecture: violation detection is reframed as learning a differentiable constraint operator rather than estimating densities or static classifiers. This positions policy as a first-class input, enabling semantic alignment, interpretability, and structural adaptivity to policy changes. Future work will address limitations including formal generalization bounds under joint distribution shift, integration of uncertainty estimation for calibrated predictions, and extension to graph-based architectures for real-world relational reasoning. INTACT establishes a foundation for policy-conditioned security monitoring where detection is explicitly guided by the constraints that define violations.
\bibliography{references}

@article{bakhshi2021anomaly,
  title={Anomaly detection in encrypted internet traffic using hybrid deep learning},
  author={Bakhshi, Taimur and Ghita, Bogdan},
  journal={Security and Communication Networks},
  volume={2021},
  number={1},
  pages={5363750},
  year={2021},
  publisher={Wiley Online Library}
}

@article{zeng2019deep,
  title={$ Deep-Full-Range $: a deep learning based network encrypted traffic classification and intrusion detection framework},
  author={Zeng, Yi and Gu, Huaxi and Wei, Wenting and Guo, Yantao},
  journal={IEEE Access},
  volume={7},
  pages={45182--45190},
  year={2019},
  publisher={IEEE}
}

@inproceedings{yu2019encrypted,
  title={An encrypted malicious traffic detection system based on neural network},
  author={Yu, Tangda and Zou, FuTai and Li, Linsen and Yi, Ping},
  booktitle={2019 international conference on cyber-enabled distributed computing and knowledge discovery (CyberC)},
  pages={62--70},
  year={2019},
  organization={IEEE}
}

@article{sattar2025anomaly,
  title={Anomaly detection in encrypted network traffic using self-supervised learning},
  author={Sattar, Sadaf and Khan, Shumaila and Khan, Muhammad Ismail and Akhmediyarova, Ainur and Mamyrbayev, Orken and Kassymova, Dinara and Oralbekova, Dina and Alimkulova, Janna},
  journal={Scientific Reports},
  volume={15},
  number={1},
  pages={26585},
  year={2025},
  publisher={Nature Publishing Group UK London}
}

@article{ji2024artificial,
  title={Artificial intelligence-based anomaly detection technology over encrypted traffic: A systematic literature review},
  author={Ji, Il Hwan and Lee, Ju Hyeon and Kang, Min Ji and Park, Woo Jin and Jeon, Seung Ho and Seo, Jung Taek},
  journal={Sensors},
  volume={24},
  number={3},
  pages={898},
  year={2024},
  publisher={MDPI}
}

@article{zhao2023ernn,
  title={ERNN: Error-resilient RNN for encrypted traffic detection towards network-induced phenomena},
  author={Zhao, Ziming and Li, Zhaoxuan and Jiang, Jialun and Yu, Fengyuan and Zhang, Fan and Xu, Congyuan and Zhao, Xinjie and Zhang, Rui and Guo, Shize},
  journal={IEEE Transactions on Dependable and Secure Computing},
  year={2023},
  publisher={IEEE}
}

@article{zhang2021dual,
  title={Dual generative adversarial networks based unknown encryption ransomware attack detection},
  author={Zhang, Xueqin and Wang, Jiyuan and Zhu, Shinan},
  journal={IEEE Access},
  volume={10},
  pages={900--913},
  year={2021},
  publisher={IEEE}
}

@inproceedings{nguyen2023deep,
  title={A deep learning anomaly detection framework with explainability and robustness},
  author={Nguyen, Manh-Dung and Bouaziz, Anis and Valdes, Valeria and Rosa Cavalli, Ana and Mallouli, Wissam and Montes De Oca, Edgardo},
  booktitle={Proceedings of the 18th International Conference on Availability, Reliability and Security},
  pages={1--7},
  year={2023}
}

@article{yang2021deep,
  title={A deep-learning-and reinforcement-learning-based system for encrypted network malicious traffic detection},
  author={Yang, Jin and Liang, Gang and Li, Beibei and Wen, Guozhu and Gao, Tianyu},
  journal={Electronics Letters},
  volume={57},
  number={9},
  pages={363--365},
  year={2021},
  publisher={Wiley Online Library}
}

@article{rahman2024explainable,
  title={Explainable anomaly detection in encrypted network traffic using data analytics},
  author={Rahman, Md Mukidur and Soumik, Md Shadman and Farids, Md Sheikh and Abdullah, Chowdhury Amin and Sutrudhar, Badhon and Ali, Mohammad and HOSSAIN, MD SHAHADAT},
  journal={Journal of Computer Science and Technology Studies},
  volume={6},
  number={1},
  pages={272--281},
  year={2024}
}

@inproceedings{canard2017blindids,
  title={BlindIDS: Market-compliant and privacy-friendly intrusion detection system over encrypted traffic},
  author={Canard, S{\'e}bastien and Diop, A{\"\i}da and Kheir, Nizar and Paindavoine, Marie and Sabt, Mohamed},
  booktitle={Proceedings of the 2017 ACM on Asia Conference on Computer and Communications Security},
  pages={561--574},
  year={2017}
}

@inproceedings{song2005formal,
  title={Formal reasoning about a specification-based intrusion detection for dynamic auto-configuration protocols in ad hoc networks},
  author={Song, Tao and Ko, Calvin and Tseng, Chinyang Henry and Balasubramanyam, Poornima and Chaudhary, Anant and Levitt, Karl N},
  booktitle={International Workshop on Formal Aspects in Security and Trust},
  pages={16--33},
  year={2005},
  organization={Springer}
}

@article{valenza2017formal,
  title={A formal approach for network security policy validation.},
  author={Valenza, Fulvio and Su, Tao and Spinoso, Serena and Lioy, Antonio and Sisto, Riccardo and Vallini, Marco and others},
  journal={J. Wirel. Mob. Networks Ubiquitous Comput. Dependable Appl.},
  volume={8},
  number={1},
  pages={79--100},
  year={2017}
}

@article{vladov2025neural,
  title={Neural Network IDS/IPS Intrusion Detection and Prevention System with Adaptive Online Training to Improve Corporate Network Cybersecurity, Evidence Recording, and Interaction with Law Enforcement Agencies},
  author={Vladov, Serhii and Vysotska, Victoria and Vashchenko, Svitlana and Bolvinov, Serhii and Glubochenko, Serhii and Repchonok, Andrii and Korniienko, Maksym and Nazarkevych, Mariia and Herasymchuk, Ruslan},
  journal={Big Data and Cognitive Computing},
  volume={9},
  number={11},
  pages={267},
  year={2025},
  publisher={MDPI}
}

@article{karaccay2020intrusion,
  title={Intrusion detection over encrypted network data},
  author={Kara{\c{c}}ay, Leyli and Sava{\c{s}}, Erkay and Alptekin, Halit},
  journal={The Computer Journal},
  volume={63},
  number={1},
  pages={604--619},
  year={2020},
  publisher={OUP}
}

@article{usman2019survey,
  title={A survey on representation learning efforts in cybersecurity domain},
  author={Usman, Muhammad and Jan, Mian Ahmad and He, Xiangjian and Chen, Jinjun},
  journal={ACM Computing Surveys (CSUR)},
  volume={52},
  number={6},
  pages={1--28},
  year={2019},
  publisher={ACM New York, NY, USA}
}

@article{karri2021security,
  title={Security and Compliance Monitoring},
  author={Karri, Nagireddy and Jangam, Sandeep Kumar},
  journal={International Journal of Emerging Trends in Computer Science and Information Technology},
  volume={2},
  number={2},
  pages={73--82},
  year={2021}
}

@article{zhang2004artificial,
  title={Artificial neural network for violation analysis},
  author={Zhang, Zhicheng and Polet, Philippe and Vanderhaegen, Fr{\'e}d{\'e}ric and Millot, Patrick},
  journal={Reliability Engineering \& System Safety},
  volume={84},
  number={1},
  pages={3--18},
  year={2004},
  publisher={Elsevier}
}

@article{nicolau2018learning,
  title={Learning neural representations for network anomaly detection},
  author={Nicolau, Miguel and McDermott, James and others},
  journal={IEEE transactions on cybernetics},
  volume={49},
  number={8},
  pages={3074--3087},
  year={2018},
  publisher={IEEE}
}

@inproceedings{zhang2012user,
  title={User intention-based traffic dependence analysis for anomaly detection},
  author={Zhang, Hao and Banick, William and Yao, Danfeng and Ramakrishnan, Naren},
  booktitle={2012 IEEE symposium on security and privacy workshops},
  pages={104--112},
  year={2012},
  organization={IEEE}
}

@article{violos2025detecting,
  title={Detecting application transitions and identifying application types for intent-based network assurance: A machine learning perspective},
  author={Violos, John and Voutsas, Fotios and Diou, Christos and Leivadeas, Aris},
  journal={Computer Networks},
  pages={111872},
  year={2025},
  publisher={Elsevier}
}

@article{gharbaoui2026assurance,
  title={Assurance and Conflict Detection in Intent-Based Networking: A Comprehensive Survey and Insights on Standards and Open-Source Tools},
  author={Gharbaoui, Molka and Sciarrone, Filippo and Fontana, Mattia and Castoldi, Piero and Martini, Barbara},
  journal={IEEE Transactions on Network and Service Management},
  volume={23},
  pages={1891--1912},
  year={2026},
  publisher={IEEE}
}

@article{izuazu2025secured,
  title={A Secured Intent-Based Networking (sIBN) with Data-Driven Time-Aware Intrusion Detection},
  author={Izuazu, Urslla Uchechi and Bensalem, Mounir and Jukan, Admela},
  journal={arXiv preprint arXiv:2511.05133},
  year={2025}
}

@inproceedings{de2024novel,
  title={A novel malicious intent detection approach in intent-based enterprise networks},
  author={De Trizio, Federica and Sciddurlo, Giancarlo and Rutigliano, Dominga and Piro, Giuseppe and Boggia, Gennaro},
  booktitle={2024 20th International Conference on Network and Service Management (CNSM)},
  pages={1--7},
  year={2024},
  organization={IEEE}
}
\bibliographystyle{ieeetr}

\appendix
\section{Appendix}
\subsection{Formal Problem Formulation}
We now present a rigorous mathematical formulation of conditional violation detection and the INTACT architecture.

\subsubsection{Conditional Violation Modeling}
Let \(x \in \mathbb{R}^d\) denote a behavioral representation of cryptographic activity. In the real dataset, \(d=7\); in the synthetic multi-intent dataset, \(d=17\). Let \(z \in \mathbb{R}^k\) encode declared security intent, where \(k=1\) for scalar lifetime thresholds or \(k>1\) for structured intent encodings. Let \(y \in \{0,1\}\) denote violation of intent \(z\).

Classical anomaly detection approximates \(g(x) \approx \mathbb{P}(y=1 \mid x)\), implicitly assuming violation is intrinsic to \(x\). This assumption fails in policy-driven systems: violation is defined relative to declared intent. We instead define the conditional violation function:
\begin{equation}
f(x,z) = \mathbb{P}(y=1 \mid x,z).
\end{equation}
This induces a family of decision manifolds:
\begin{equation}
\mathcal{M}(z) = \{x \in \mathbb{R}^d : f(x,z)=0.5\},
\end{equation}
representing a policy-parameterized manifold in \(\mathbb{R}^{d+1}\) rather than a single separating surface.

For scalar threshold intents \(z = \tau\), suppose an unknown latent violation score \(g(x)\) satisfies \(y = \mathbf{1}(g(x) > \tau)\). Then \(\mathbb{P}(y=1 \mid x,\tau) = \mathbf{1}(g(x) > \tau)\). INTACT approximates a differentiable relaxation:
\begin{equation}
f(x,\tau) = \sigma\big(g_\theta(x) - h_\theta(\tau)\big),
\end{equation}
where \(g_\theta\) and \(h_\theta\) are learned nonlinear mappings—a learnable conditional comparator generalizing pure thresholding.

\subsubsection{Factorized Representation and Interaction}
We parameterize \(f\) through factorization:
\begin{equation}
f(x,z) = \sigma\!\left( \phi\big(\psi_b(x), \psi_z(z)\big) \right),
\end{equation}
with \(\psi_b : \mathbb{R}^d \to \mathbb{R}^m\), \(\psi_z : \mathbb{R}^k \to \mathbb{R}^n\), and \(\phi : \mathbb{R}^{m+n} \to \mathbb{R}\). This enforces decomposition \(\mathbb{R}^{d+k} = \mathbb{R}^d \oplus \mathbb{R}^k\) followed by nonlinear coupling.

\paragraph{Behavioral Encoder.}
\begin{equation}
\psi_b(x) = W_3 \rho\!\left( W_2 \rho(W_1 x + b_1) + b_2 \right) + b_3,
\end{equation}
with ReLU activation \(\rho\), defining \(\psi_b : \mathbb{R}^d \to \mathcal{H}_b \subset \mathbb{R}^m\). By universal approximation, for compact \(\Omega \subset \mathbb{R}^d\), any continuous violation score can be approximated arbitrarily closely.

\paragraph{Intent Encoder.}
\begin{equation}
\begin{aligned}
\psi_z(z) = & V_2 \rho(V_1 z + c_1) \\
            & + c_2,
\end{aligned}
\end{equation}
defining \(\psi_z : \mathbb{R}^k \to \mathcal{H}_z \subset \mathbb{R}^n\). For scalar \(z\), this learns a nonlinear basis expansion of policy magnitude, enabling smooth policy interpolation: \(\| z_1 - z_2 \| \to 0 \Rightarrow \| \psi_z(z_1) - \psi_z(z_2) \| \to 0\). Decision boundaries thus deform continuously with policy.

\paragraph{Fusion.}
Let \(h_b = \psi_b(x)\), \(h_z = \psi_z(z)\). Concatenation yields \(h = [h_b ; h_z] \in \mathbb{R}^{m+n}\). The decision layer computes:
\begin{equation}
\phi(h) = u^\top \rho(W_4 h + b_4) + b_5,
\end{equation}
with final output \(f(x,z) = \sigma(\phi(h))\). Hidden nonlinearities enable interaction terms \(x_i z_j\), \(x_i x_j z_k\), and higher-order compositions, supporting threshold comparisons, intent-conditioned feature reweighting, and cross-trace relational inference.

\subsubsection{Optimization and Policy-Dependent Gradient Flow}
The empirical risk minimizes binary cross-entropy:
\begin{equation}
\begin{split}
\mathcal{L}(\theta) = & -\frac{1}{N} \sum_{i=1}^N \Big[ y_i \log f_\theta(x_i,z_i) \\
& + (1-y_i)\log(1-f_\theta(x_i,z_i)) \Big].
\end{split}
\end{equation}
The gradient decomposes as:
\begin{equation}
\nabla_\theta \mathcal{L} = \frac{1}{N} \sum_{i=1}^N (f_\theta(x_i,z_i) - y_i) \nabla_\theta \phi_\theta(x_i,z_i),
\end{equation}
where \(\phi_\theta(x_i,z_i) = \phi(\psi_b(x_i), \psi_z(z_i))\). Applying the chain rule:
\begin{equation}
\nabla_\theta \phi = \frac{\partial \phi}{\partial h_b} \nabla_\theta \psi_b + \frac{\partial \phi}{\partial h_z} \nabla_\theta \psi_z.
\end{equation}
Policy encoding thus directly influences parameter updates in both encoders; representation learning is jointly optimized with constraint semantics.

\subsubsection{Geometric Interpretation}
Classical models learn a fixed \(\mathcal{M} \subset \mathbb{R}^d\). INTACT learns \(\mathcal{M} \subset \mathbb{R}^{d+k}\), with projection \(\mathcal{M}(z) = \{x : f(x,z)=0.5\}\)—a hypersurface family indexed by policy. Under regularity conditions:
\begin{equation}
\frac{\partial f}{\partial z} = \sigma'(\phi) \frac{\partial \phi}{\partial h_z} \frac{\partial \psi_z}{\partial z},
\end{equation}
explicitly showing decision boundary sensitivity to policy. INTACT implements a differentiable comparator between behavioral score and policy embedding.

\subsubsection{Hypothesis Class and Complexity}
Define \(\mathcal{F} = \{ f_\theta(x,z) \}\). Without factorization, VC-dimension scales with \(d+k\). With factorization \(f(x,z) = \phi(\psi_b(x),\psi_z(z))\), capacity decomposes additively rather than multiplicatively. Effective complexity scales as \(\mathcal{O}(m + n)\) with \(m \ll d\), \(n \ll k\), enabling parameter sharing across intents and improved sample efficiency.

\subsection{Theoretical Reformulation: Conditional Constraint Learning}
INTACT reformulates violation detection as policy-parameterized risk minimization—a fundamental shift from classical statistical learning.


\subsubsection{From Static to Conditional Risk}
Classical supervised learning solves \(\min_{\theta} \mathbb{E}_{(x,y) \sim \mathcal{D}} \ell(g_\theta(x), y)\), assuming labeling function \(y = h(x)\) intrinsic to \(x\). In cryptographic monitoring, \(y = h(x,z)\) depends on declared intent. The Bayes-optimal rule becomes:
\begin{equation}
f^*(x,z) = \mathbb{P}(y=1 \mid x,z).
\end{equation}
INTACT solves:
\begin{equation}
\min_{\theta} \mathbb{E}_{(x,z,y) \sim \mathcal{D}} \ell(f_\theta(x,z), y),
\end{equation}
replacing \(\mathcal{D}(x,y)\) with \(\mathcal{D}(x,z,y)\). The label distribution now depends jointly on behavior and declared constraint.

\subsubsection{Violation as Constraint Satisfaction}
For scalar threshold \(z = \tau\), violations arise from \(y = \mathbf{1}(g(x) > \tau)\). Classical classifiers learn \(g(x)\) implicitly. INTACT learns:
\begin{equation}
f(x,\tau) = \sigma(g_\theta(x) - h_\theta(\tau)),
\end{equation}
a differentiable operator \(\mathcal{C}_\theta : (x,\tau) \mapsto \text{constraint violation}\)—a parameterized constraint comparator rather than a static decision boundary.

\subsubsection{Structured Hypothesis Space Enlargement}
Classical: \(\mathcal{G} = \{ g_\theta : \mathbb{R}^d \to [0,1] \}\). INTACT: \(\mathcal{F} = \{ f_\theta : \mathbb{R}^{d+k} \to [0,1] \}\) with factorization \(f_\theta(x,z) = \phi(\psi_b(x), \psi_z(z))\). While the input space expands, effective complexity remains controlled:
\begin{itemize}
    \item Behavioral representation \(m \ll d\)
    \item Policy representation \(n \ll k\)
    \item Effective capacity \(\mathcal{O}(m + n)\) rather than \(\mathcal{O}(d+k)\)
\end{itemize}
This structured enlargement enables policy adaptivity without uncontrolled VC-dimension growth.

\subsubsection{Policy-Parameterized Decision Manifolds}
Traditional classifiers: fixed \(\mathcal{M} \subset \mathbb{R}^d\). INTACT: smooth family \(\mathcal{M}(z) = \{x \mid f(x,z)=0.5\}\) with boundary sensitivity:
\begin{equation}
\frac{\partial \mathcal{M}}{\partial z} = - \frac{\partial f/\partial z} {\|\nabla_x f\|}.
\end{equation}
The model learns a policy-indexed geometric object in feature space, not merely a classifier.

\subsubsection{Connection to Lagrangian Relaxation}
Hard constraints \(\min_x \mathcal{L}(x) \; \text{s.t.} \; c(x) \le 0\) relax to \(\mathcal{L}(x) + \lambda c(x)\). INTACT learns a data-driven analog where:
\begin{itemize}
    \item \(z\) acts as constraint parameter
    \item \(\psi_z(z)\) modulates violation severity
    \item Sigmoid output approximates constraint satisfaction
\end{itemize}
The model implicitly learns a differentiable surrogate for constraint enforcement.

\subsubsection{Distinction from Density-Based Detection}
Density-based methods estimate \(p(x)\) and flag low-density regions. This fails when:
\begin{itemize}
    \item Violations lie in high-density regions (e.g., downgrade events)
    \item Rare but compliant behavior should not be flagged
\end{itemize}
INTACT approximates \(\mathbb{P}(y=1 \mid x,z)\)—a semantic, frequency-independent quantity. Anomaly detection is replaced by policy-conditioned violation inference.

\section{Advantages Over Existing Approaches}

\paragraph{Explicit Policy Alignment.} Classical anomaly detectors (Isolation Forest, Deep SVDD, autoencoders) implicitly approximate decision boundaries from data density, assuming violations correspond to statistical rarity. In contrast, INTACT directly conditions on the declared policy parameter, aligning detection with operational semantics rather than frequency. For lifetime detection, this yields near-deterministic separability while maintaining generalization under covariate shift—a capability density-based methods lack.

\paragraph{Multi-Intent Unification.} Traditional supervised methods require independent classifiers for each violation type (reuse, downgrade, lifetime), leading to redundant parameters and fragmented representations. INTACT supports a unified backbone that dynamically adapts via intent embeddings, enabling shared feature extraction while preserving intent-specific decision boundaries—reducing complexity and improving sample efficiency.

\paragraph{Relational Violation Modeling.} Unsupervised reconstruction and hypersphere-based methods excel at magnitude anomalies but fail on relational violations like key reuse, which require cross-trace reasoning. By conditioning on intent and learning discriminative representations, INTACT captures these relational dependencies effectively, addressing a fundamental blind spot in existing approaches.

\paragraph{Robustness Under Distribution Shift.} Reconstruction-based models exhibit inflated AUROC under covariate scaling but suffer calibration collapse (low F1). INTACT's explicit intent encoding preserves discrimination under moderate shift, and while threshold metrics degrade, the model's structural capacity for policy adaptation exceeds static classifiers that must be retrained for new regimes.
\end{document}